\documentstyle[12pt]{article}

\begin{document}
{\sf\begin{center}\noindent {\Large\bf Conformal maps in periodic dynamo flows and in stretch-twist suppression on Riemannian manifolds}\\[3mm]

by \\[0.3cm]

{\sl L.C. Garcia de Andrade}\\

\vspace{0.4cm} Departamento de F\'{\i}sica
Te\'orica -- IF -- Universidade do Estado do Rio de Janeiro-UERJ\\[-3mm]
Rua S\~ao Francisco Xavier, 524\\[-3mm]
Cep 20550-003, Maracan\~a, Rio de Janeiro, RJ, Brasil\\[-3mm]
Electronic mail address: garcia@dft.if.uerj.br\\[-3mm]
\vspace{1cm} {\bf Abstract}
\end{center}
\paragraph*{}
Examples of conformal dynamo maps have been presented earlier [Phys
Plasmas \textbf{14}(2007)] where fast dynamos in twisted magnetic
flux tubes in Riemannian manifolds were obtained. This paper shows
that conformal maps, under the Floquet condition, leads to
coincidence between exponential stretching or Lyapunov exponent,
conformal factor of fast dynamos. Unfolding conformal dynamo maps
can be obtained in Riemann-flat manifolds since here, Riemann
curvature plays the role of folding. Previously, Oseledts [Geophys
Astrophys Fluid Dyn \textbf{73} (1993)] has shown that the number of
twisted and untwisting orbits in a two torus on a compact Riemannian
manifold induces a growth of fast dynamo action. In this paper, the
stretching of conformal thin magnetic flux tubes is constrained to
vanish, in order to obtain the conformal factor for non-stretching
non-dynamos. Since thin flux tube can be considered as a twisted or
untwisting two-torus map, it is shown that the untwisting, weakly
torsion, and non-stretching conformal torus map cannot support a
fast dynamo action, a marginal dynamo being obtained. This is an
example of an anti-fast dynamo theorem besides the ones given by
Vishik and Klapper and Young [Comm Math Phys \textbf{173}(1996)] in
ideally high conductive flow. From the Riemann curvature tensor it
is shown that new conformal non-dynamo, is actually singular as one
approaches the magnetic flux tube axis. Thus conformal map
suppresses the stretching directions and twist, leading to the
absence of fast dynamo action while Riemann-flat unfolding manifolds
favors non-fast dynamos.{\textbf{Key-words}:anti-dynamo theorems,
conformal stretching.\bf PACS
numbers:\hfill\parbox[t]{13.5cm}{2.40.Hw:differential geometries.
91.25.Cw-dynamo theories.}}

\newpage
 \section{Introduction}
 An anti-dynamo theorem by Zeldovich \cite{1}, shows that planar flows cannot support
 dynamo action. Together with Cowling \cite{2} theorem they compose the two first anti-dynamo theorems in magnetohydrodynamics (MHD). As it is well-known,
 the stretching is a fundamental ingredient in fast dynamo \cite{3}
 action and leads to exponential stretching. This important role played by stretching in the building of dynamo action led Vishik \cite{4} to formulate the
 first anti-fast-dynamo theorem, which states that no fast dynamo action can be obtained in non-stretching flows, and only fast dynamos could be obtained.
 More recently Klapper and Young \cite{5} have extended the $\cal{C}^{\infty}$-differentiable anti-fast-dynamo theorem to ${\cal{C}}^{2}$. In this´paper, following the conformal
 technique transformation path, so much used in Einstein's general theory pseudo-Riemannian spacetime ,a conformal dynamo flow is proposed by considering a
 generalization of the Ricca's flux tube \cite{6} to the conformal Riemann metric to his thin twisted magnetic flux tube. An interesting aspect of this
 technique applied to dynamo theory, is that contrary to what happens in
 Einstein's theory of gravity, here the conformally flat metric is
 not flat, in the sense that the Riemann curvature tensor components do not all vanish. As
 it is shown in the paper, Riemann curvature components depend upon the Frenet curvature and torsion of the magnetic flux tube axis. Throughout the paper, to
 simplify computations the tube is consider as helical \cite{7} which is so commonly used in plasma astrophysics. This implies also that folding plays the
 role of Riemann curvature and folding in this sense is not always destructive in the anti-dynamo case considered here since, all components of the
 Riemann curvature tensor are positive and the folding is constructive \cite{3}. In this way though there is no cancellation of magnetic fields,
 presence of non-stretching even in the twisting case, leads to a slow dynamo or non-dynamo at all. Actually in the first part of the paper an anti-fast
 dynamo like theorem is obtained forcing the stretching term in non-dynamo equation to vanish. This is mainly due to the fact that, cancellation of
 magnetic fields are possible in regions of folding or curvature. In tubes, the Riemann curvature concentrates strongly the magnetic fields inside the
 tubes in order that the cancellation cannot take place. In the second part of the paper one considers that actually that conditions and constrations
 that led to the non-dynamo are actually too restrictive or demanding and that fast conformal dynamos \cite{8} can be obtained in Riemannian spaces.
 Actually Arnold, Zeldovich, Ruzmaikin and Sokoloff \cite{9} and Chicone
 and Latushkin \cite{10} have previously shown that uniform stretching in flows could lead to fast dynamos in compact Riemannian manifolds.
 Though in the first part of the paper one addresses the conformal slow dynamos in the zero-resistivity limit, in the second part one considers
 conformal fast dynamos in diffusive media. Actually diffusion processes in Riemannian manifolds have been already considered
 by S. Molchanov \cite{11}, without addressing the problem of dynamo action. Indeed Soward \cite{12},
 argued earlier that fast dynamo actions would still be possible in
 regions where no non-stretching flows would be presented, such as in some curved
 surfaces. Here one argues that, there is no reason to believe that these surfaces could not be Riemannian. In this paper the well-known stretch,twist and
 fold method developed by Vainshtein and Zeldovich \cite{13} seems to be effective due to the conformal dynamo map. Actually from the mathematical point
 of view, Osedelets \cite{14} has shown that by investigating the fast dynamo Lyapunov exponent on a two-torus map, is the untwisting and twisting number. To resume, due to the frozen magnetic
 field from the zero resistivity hypothesis, the flow stretching necessary leads to the geometrical stretching of the Riemann metric. It seems that similar to
 Soward's arguement, the Riemann curved surfaces of the tube which are non-stretched, necessarily leads to non-dynamos, dynamo solutions being possible
 off the surfaces. The paper is
 organized as
 follows: In section II the conformal map is shown to lead , under Floquet condition, to the fast dynamo where the conformal Riemann metric factor is
 the exponential stretching with Lyapunov exponent of chaotic flows. In section III an anti-fast dynamo is proved and the Riemann curvature tensor are computed and stretching instability of
 the conformal class discussed in section II is obtained. In section
 III the suppression of stretching and twist by conformal action is shown to lead to a marginal, non-fast-dynamo in compact Riemannian manifold.
 In section IV a conformal dynamo map in twisted magnetic flux tube is obtained. Discussions and future prospects are presented in section V.
\newpage
\section{Conformal fast dynamo maps on periodic flows}
Childress and Gilbert \cite{3} have discussed in detail the
suppression of stretching and twist on the flux tubes. In this
section a very simple proof is done of the fact that the conformal
metric factor is given by the exponential stretching leading to the
conformal fast dynamo. In the next two sections, suppression of
stretch and twist in turn, is showed to be connect to non-fast
dynamos, providing us with a new application of Vishik´s and Klapper
and Young anti-fast dynamo theorems. In this section one shall
basically deal with Cartesian coordinates in three-dimensional
Euclidean $\textbf{E}^{3}$. Let us start by considering the
conformal transformation or rescaling ${\Omega}$, in the Cartesian
coordinates as
\begin{equation}
{x'}^{i}={\Omega}^{-1}x^{i} \label{1}
\end{equation}
where the indices $(i,j=1,2,3)$. From this equations
\begin{equation}
\frac{{\partial}{x'}^{i}}{{\partial}x^{j}}={\Omega}^{-1}{{\delta}^{i}}_{j}
\label{2}
\end{equation}
Thus from the invariance of line elements of Riemann differential
geometry ,and by the law of transformation of the Riemann metric
components $g_{ij}$ one has
\begin{equation}
{g'}_{ij}={\Omega}^{2}(\textbf{x},t)g_{ij} \label{3}
\end{equation}
By considering the general expression of the magnetic field
variation under the general coordinate transformation
$\textbf{x'}\rightarrow{\textbf{x}}$, one has to use the Jacobian
transformation \cite{3}
\begin{equation}
{B'}^{i}={\textbf{J}^{i}}_{j}{B}^{j} \label{4}
\end{equation}
where the Jacobian matrix $\textbf{J}^{ij}$ is
\begin{equation}
{{\textbf{J}}^{i}}_{j}:=\frac{{\partial}{x'}^{i}}{{\partial}x^{j}}
\label{5}
\end{equation}
Application of these formulas into the conformal transformation of
the partial derivative in the gradient
\begin{equation}
({\nabla}')_{i}:={{\partial}'}_{i}={\Omega}{{\nabla}}_{i} \label{6}
\end{equation}
comparision with expression (\ref{1}) helps us to determine the
conformal factor ${\Omega}$ as
\begin{equation}
{\Omega}^{-1}{{\delta}^{i}}_{j}={\textbf{J}^{i}}_{j} \label{7}
\end{equation}
which from (\ref{4}) yields
\begin{equation}
{B'}_{i}={\Omega}{B}_{i} \label{8}
\end{equation}
when one consider the periodic flows under the Flouquet condition
\begin{equation}
{B'}_{i}(\textbf{x},t+T)=e^{{\gamma}({\eta})T}B_{i}(\textbf{x},t)\label{9}
\end{equation}
which from (\ref{8}) yields
\begin{equation}
{\Omega}=e^{{\gamma}T} \label{10}
\end{equation}
where ${\gamma}$ is the growth of magnetic field. Therefore one must
conclude that under certain conditions on diffusion, the conformal
factor can be associated with exponential chaotic stretching. In the
next section one adopts the more simple assumption of
zero-resistivity plasmas. Since the dynamo action is basically
discovered by solving the self-induction equation, most of the times
in Cartesian coordinates, now let us investigate this equation in
this coordinates as
\begin{equation}
d_{t}{B'}^{i}={B'}^{j}{{\partial}'}_{j}v^{i}+{\eta}{{{\partial}'}^{k}{{\partial}'}_{k}}B^{i}\label{11}
\end{equation}
Now by considering the zero-resistivity ideal plasma, where
resistivity ${\eta}$ vanishes, one has
\begin{equation}
d_{t}{B'}^{i}={B'}^{j}{{\partial}'}_{j}{v'}^{i}\label{12}
\end{equation}
By taking into account that flow vector transform conformally as
\begin{equation}
{v'}_{i}={\Omega}v_{i}\label{13}
\end{equation}
one obtains the following constraint on the omega conformal factor
as
\begin{equation}
d_{t}{\Omega}+\frac{1}{3}({\textbf{v}}.{\nabla}){\Omega}=0\label{14}
\end{equation}
where one has made use of the unprimed self-induced equation to
achieved this equation. This equation is very important even when
periodic flows are absent, as we shall see by the end of this
section , when the unfolding is connected with the vanishing of the
Riemann tensor, or like in geometers jargon, a Riemann-flat
manifold. In this case one shall drop the Cartesian coordinates used
here and consider the Riemannian tubular coordinates to a, in
principle curved Riemannian flux tube. Since we are considering in
this section all conformal dynamo maps and leave the anti-dynamo
theorems and non-dynamos to the last two sections, let us consider
first the case of flux tubes where the conformal factor ${\Omega}$
depends only on the radial coordinate-r. Thus by expression
\begin{equation}
({\textbf{v}}.{\nabla}){\Omega}=0\label{15}
\end{equation}
applied to magnetic flux tube coordinates $(r,{\theta}_{R},s)$,
yields
\begin{equation}
({\textbf{v}}.{\nabla}){\Omega}=v^{r}{\partial}_{r}{\Omega}(r)=0\label{16}
\end{equation}
which yields $v^{r}=0$ or $v_{r}=0$ since the Riemann metric of the
tube ((see formulas in section II)), is diagonal. Thus this choice
of conformal factor is consistent with the usual choice of flux
tubes, where the flow is confined inside the tube. This is also the
approach in plasma torus called tokamaks by plasma physicists.
Actually since the folding in flux tubes maybe be represented by the
Riemann curvature tensor, constructive folding that leads to fast
dynamos, can be obtained by the vanishing of folding or by vanishing
of the Riemann curvature tensor. By the formula, for the Riemann
curvature tensor for ${\Omega}(r)$ below , one finds that
$R_{r{\theta}r{\theta}}:=0$ constraint reduces to a differential
equation for the conformal factor as
\begin{equation}
r({\Omega}^{2})"+({\Omega}^{2})'-\frac{1}{2{\Omega}^{2}}({\Omega}^{2})'r=0\label{17}
\end{equation}
which has a very simple particular solution compatible with
(\ref{14}) which is ${\Omega}(r)=r$, which yields the fast dynamo
conformal thin flux tube Riemannian metric as
\begin{equation}
d{s_{0}}^{2}={r}^{2}[dr^{2}+r^{2}d{\theta}^{2}]+ds^{2}\label{18}
\end{equation}
In this way a fast conformal dynamo map was obtained in periodic and
plasma flows inside Riemann-flat thin tubes. Actually from the
incompressible flows condition
\begin{equation}
{\nabla}.{\textbf{v}}=0\label{19}
\end{equation}
and the conformal invariance, one is led to conclude that the
equation (\ref{15}), is nothing but the statement that the conformal
factor does not change along the streamlines. The same is true to
the magnetic field lines which solenoidal vector condition
\begin{equation}
{\nabla}.{\textbf{B}}=0\label{20}
\end{equation}
also leads to
\begin{equation}
({\textbf{B}}.{\nabla}){\Omega}=0\label{21}
\end{equation}
which is relation which expresses the fact that the conformal factor
does not vary along magnetic field lines. The fact that these two
relations are similar, stems from the fact that the plasma flow is
ideal or resistivity free, and the frozen condition which tights the
magnetic field lines to the streamlines is given. Note also that
magnetic expression, the case of magnetic flux tubes yields that
$B^{r}=0$ or that the radial component of the magnetic field
vanishes due to the confinement condition on astrophysical plasma
flux tubes and plasma torus devices.
\newpage

\section{Conformal non-stretching marginal tube dynamos} This
section contains the presentation and a simple proof of the
following theorem:
\newline{\textbf{Theorem 1:}} Let us consider a Riemannian compact and curved conformal thin magnetic twisted flux
tube $\cal{T}$ embbeded in the Euclidean space $\textbf{E}^{3}$.
Thus a class of conformal ${\cal{C}}^{1} $-differentiable, slow
dynamos, in diffusive-free manifolds can be obtained by imposing the
constraint of vanishing of the stretching term of self-induction
magnetic equation. Of course, when one considers the resistivity in
dynamo flows, the Klapper-Young's ${\cal{C}}^{2}$-differentiable
condition is needed.\newline \textbf{Proof:} By considering the
conformal metric to the Riemann-flat thin magnetic flux tube metric
\begin{equation}
{ds_{0}}^{2}=dr^{2}+r^{2}d{{\theta}_{R}}^{2}+ds^{2} \label{22}
\end{equation}
as
\begin{equation}
dl^{2}:={\Omega}^{2}(r,s){ds_{0}}^{2}={\Omega}^{2}[dr^{2}+r^{2}d{{\theta}_{R}}^{2}+ds^{2}]
\label{23}
\end{equation}
where the coordinates are $(r,{\theta}_{R},s)$ and
${\theta}(s)={\theta}_{R}-\int{{\tau}(s)ds}$, ${\tau}$ being the
Frenet torsion. The conformal factor is given by the real function
${\Omega}$ which here we must consider at least ${\cal{C}}^{1}$
differentiable. Now let us consider the self-induction equations as
\begin{equation}
d_{t}\textbf{B}=(\textbf{B}.{\nabla})\textbf{v}+{\eta}{\nabla}^{2}\textbf{B}
\label{24}
\end{equation}
where the resistivity ${\eta}=0$,since one is assuming the ideal
plasma case. By considering that the magnetic field is strictly
confined along and inside the tube, and that the resistivity free
case leads to the frozen condition , the flow is also confined on
the tube which allows us not to stretch the magnetic flux tube
without stretching the magnetic field. Therefore,as one shall
demonstrate in this section the non-stretching of the magnetic flow
along the tube implies necessarily a geometrical non-stretching
along the magnetic field axis direction of the the flux tube. The
above confinement hypothesis leads to the vanishing of radial
components of magnetic field and flow, thus the stretching term in
the conformal metric yields
\begin{equation}
(\textbf{B}.{\nabla})\textbf{v}=\textbf{e}_{r}(\frac{{\partial}_{s}{\Omega}}{\Omega})A(r,s)+
\textbf{e}_{\theta}[\frac{{B}_{\theta}}{{\Omega}r}C(r,s)+
\frac{{B}_{s}}{\Omega}({{\tau}_{0}}^{-1}\frac{{\partial}_{s}{\Omega}}{\Omega}v_{s}+{\partial}_{s}v_{\theta})]-\textbf{t}[\frac{{B}_{\theta}}{{\Omega}}(v_{\theta}\frac{{\partial}_{s}{\Omega}}{\Omega})+\frac{B_{s}}{\Omega}{{\tau}_{0}}^{-1}v_{\theta}\frac{{\partial}_{s}{\Omega}}{\Omega}]
\label{25}
\end{equation}
where
\begin{equation}
A(r,s):=B_{\theta}{{\tau}_{0}}^{-1}v_{\theta}-B_{s}v_{s} \label{26}
\end{equation}
and
\begin{equation}
C(r,s):=-{{\tau}_{0}}^{-1}{\partial}_{s}v_{\theta}+rv_{s}\frac{{\partial}_{s}{\Omega}}{\Omega}
\label{27}
\end{equation}
where by assuming the vanishing of stretching expression (\ref{27})
and considering the other differential equation from the solenoidal
vector condition
\begin{equation}
{\nabla}.\textbf{B}=[B_{\theta}{{\tau}_{0}}^{-1}-rB_{s}]{\partial}_{s}ln{\Omega}+{\Omega}^{2}{{\tau}_{0}}^{-1}{\partial}_{s}B_{\theta}=0
\label{28}
\end{equation}
A almost trivial solution of this system of PDEs is given simply by
\begin{equation}
\frac{{\partial}_{s}{\Omega}}{\Omega}=0 \label{29}
\end{equation}
and
\begin{equation}
{\partial}_{s}B_{\theta}=0 \label{30}
\end{equation}
This last expression implies that the tube is untwisting.
${\Box}$\newline
 This simple case and actually all solutions of the
self-induction condition for this metric possesses a problem which
is the fact that the metric that expands along both directions of
the tube, namely the tube cross-section and the toroidal direction
along the tube magnetic axis, certainly "explodes" and shall never
be a dynamo at all, from the very beginning. To avoid these kind of
pathologies, one makes use of the non-uniform metric
\begin{equation}
{ds_{0}}^{2}={\Omega}^{2}[dr^{2}+r^{2}d{{\theta}_{R}}^{2}]+K^{2}ds^{2}
\label{31}
\end{equation}
where when the $K$ is given as above represents a flux tube. Thus by
computing the same stretching term above with the new sectionally or
piecewise metric (\ref{31}) yields
\begin{equation}
\frac{{\partial}_{r}(r{\Omega})}{{\Omega}^{2}}{B}_{\theta}v_{\theta}=-\frac{B_{s}}{{\Omega}K}v_{s}{\partial}_{r}{K}
\label{32}
\end{equation}
\begin{equation}
\frac{{B}_{\theta}}{{\Omega}K}{\partial}_{s}{\Omega}v_{s}=-\frac{B_{s}}{K}({\partial}_{s}B_{\theta})+\frac{B_{s}}{r{\Omega}}{{\tau}_{0}}^{-1}
{\partial}_{s}K \label{33}
\end{equation}
\begin{equation}
\frac{{B}_{\theta}}{\Omega}(\frac{v_{\theta}}{K})=-\frac{{B}_{s}}{K}[{\partial}_{r}K-{{\tau}_{0}}^{-1}\frac{v_{s}}{\Omega}{\partial}_{s}K]
\label{34}
\end{equation}
The solenoidal condition for the magnetic fields is
\begin{equation}
({B}_{\theta}-{\tau}_{0}{B}_{s}){\partial}_{s}ln(K{\Omega})={\partial}_{s}B_{\theta}
\label{35}
\end{equation}
From these expressions one may be able to prove the following
theorem.\newline \textbf{Theorem \textbf{2}}: A piecewise conformal
flux tube geometry leads to a non-stretching tube, obtained from a
non-stretching flow. This is a particular application of Vishik's
anti-fast dynamo theorem.
\newline
\textbf{Proof}: Again untwisting solution
${\partial}_{s}B_{\theta}=0$, now reduces the equations (\ref{16})
through (\ref{19}) leads to
\begin{equation}
{\partial}_{s}K= 0 \label{36}
\end{equation}
which represents the non-stretching along the toroidal direction,
and
\begin{equation}
{\partial}_{r}K= 0 \label{37}
\end{equation}
from (\ref{20}). These results shows that the tube is geometrically
non-stretching along the toroidal and poloidal directions.${\Box}$ A
corollary can be obtained from this theorem as:\newline

\textbf{Corollary}: The Riemannian metric of the conformal
non-dynamo map is given by:
\begin{equation}
{ds_{0}}^{2}={(\frac{{\Omega}_{0}}{r})}^{2}[dr^{2}+r^{2}d{{\theta}_{R}}^{2}]+{K_{0}}^{2}ds^{2}
\label{38}
\end{equation}
where $K_{0}$ is constant. As we shall see below ,when $K_{0}=1$ one
is left with a conformal non-stretched thin magnetic flux tube.
\newline
\textbf{Proof}: From the expression (\ref{16}) one obtains
\begin{equation}
{\partial}_{r}({\Omega}r)=0 \label{39}
\end{equation}
which yields ${\Omega}=\frac{{\Omega}_{0}}{r}$ and the conformal
Riemann metric (\ref{22}). In the next section, one shall compute
the effects on Riemann curvature folding in the conformal metric.
\newpage
\section{Curvature-folding and suppression of stretch-twist in conformal marginal dynamos}
 Since the Riemann curvature is not in
principle connected to the stretching of the tube but to its
folding, one is able to compute the curvature of the tube as Let us
now consider the Riemann metric of a flux tube in curvilinear
coordinates $(r,{\theta}_{R},s)$, where
${\theta}={\theta}_{R}-\int{{\tau}(s)ds}$. Here, accordingly to
Ricca's flux tube metric, $K^{2}=(1-{\kappa}rcos{\theta})^{2}$. This
expression contributes to the Riemann curvature components. These
can be easily computed with the tensor computer package in terms of
Riemann components. Adapting the general relativity (GR) tensor
package to three-dimensions, yields the following curvature
components expressions
\begin{equation}
R_{1313}=R_{rsrs}=
-\frac{1}{4K^{2}}[2K^{2}{\partial}_{r}D(r,s)-D^{2}]=-\frac{1}{2}\frac{K^{4}}{r^{2}}=-\frac{1}{2}r^{2}{\kappa}^{4}cos^{2}{\theta}
\label{40}
\end{equation}
\begin{equation}
R_{2323}=R_{{\theta}s{\theta}s}= -\frac{r}{2}D(r,s)=
-{K^{2}}\label{41}
\end{equation}
where $D:={\partial}_{r}K^{2}$. For thin tubes,
$K^{2}(r,s)\approx{1}$ $(r\approx{0})$, these Riemann curvature
components reduce to
\begin{equation}
R_{1313}=R_{rsrs}= -\frac{1}{r^{2}} \label{42}
\end{equation}
Let us now compute the Riemann curvature components of the
\begin{equation}
R_{1313}=R_{rsrs}=
-\frac{1}{4K^{2}}[2K^{2}{\partial}_{r}D(r,s)-D^{2}]=-\frac{1}{2}\frac{K^{4}}{r^{2}}=-\frac{1}{2}r^{2}{\kappa}^{4}cos^{2}{\theta}
\label{43}
\end{equation}
Now let us compute the Riemann tensor for metric (\ref{22}). Since
now most of the Riemann components involve the derivatives with
respect to toroidal coordinate-s and derivatives of constant
$K`_{0}$, the only non-vanishing component of the Riemann metric is
\begin{equation}
R_{1212}=R_{r{\theta}r{\theta}}=
-\frac{r}{4}[r({\Omega}^{2})"+({\Omega}^{2})'-\frac{1}{2{\Omega}^{2}}({\Omega}^{2})'r]\label{44}
\end{equation}
which from (\ref{22}) yields
\begin{equation}
R_{1212}= -\frac{3{{\Omega}_{0}}^{2}}{r^{2}} \label{45}
\end{equation}
Thus one notices that the Riemann curvature is singular along the
non-dynamo flux tube magnetic axis. Now let us prove the following
theorem, which is a stronger version of the previous anti-dynamo
theorem for flux tubes:
\newline
\textbf{Theorem 3:} The non-stretching, untwisted conformal dynamo
two-torus map on a compact Riemannian manifold, is unstable and
leads to a marginal non-fast dynamo, along the Riemann metric of
conformally thin untwisted magnetic flux tube. This can be
considered as a suppression of the stretching directions by
conformal action.
\newline
\textbf{Proof:} Let us now analyse the LHS of the self-induction
magnetic equation
\begin{equation}
d_{t}\textbf{B}=
d_{t}(h^{2}B_{\theta}\textbf{e}_{\theta}+h^{3}B_{s}\textbf{t}))=d_{t}[(r^{-1}{\Omega}^{-2}B_{\theta}\dot{\Omega}+{r^{-1}}{\Omega}^{-1}\dot{B_{\theta}})\textbf{e}_{\theta}
+{r^{-1}}{\Omega}^{-1}\dot{B_{\theta}}\dot{\textbf{e}_{\theta}}-B_{s}\dot{\textbf{t}}+\dot{B_{s}}{\textbf{t}}]\label{46}
\end{equation}
making use of the following definitions of the poloidal and toroidal
components of the magnetic fields as
\begin{equation}
B_{\theta}={B^{0}}_{\theta}e^{{\gamma}t}\label{47}
\end{equation}
and
\begin{equation}
B_{s}={B^{0}}_{s}e^{{\gamma}t}\label{48}
\end{equation}
where superscript-0 index refers to stationary quantities. By making
use of the identities on the relations between the two frames , of
tube and Frenet frame
\begin{equation}
\dot{\textbf{e}_{\theta}}=-{\omega}_{0}\textbf{e}_{r}-sin{\theta}{{\tau}_{0}}^{2}{\textbf{t}}\label{49}
\end{equation}
in the helical tube. This computation yields three scalar equations
along the tube frame
$(\textbf{e}_{r},\textbf{e}_{\theta},\textbf{t})$ transformed to the
Frenet frame with the help of equations
\begin{equation}
\dot{\textbf{t}}=[{\kappa}'\textbf{b}-{\kappa}{\tau}\textbf{n}]
\label{50}
\end{equation}
\begin{equation}
\dot{\textbf{n}}={\kappa}\tau\textbf{t} \label{51}
\end{equation}
\begin{equation}
\dot{\textbf{b}}=-{\kappa}' \textbf{t} \label{52}
\end{equation}
Together with the flow derivative
\begin{equation}
\dot{\textbf{t}}={\partial}_{t}\textbf{t}+(\vec{v}.{\nabla})\textbf{t}
\label{53}
\end{equation}
one obtains
\begin{equation}
{({\Omega}r)}^{-1}[(\frac{\dot{\Omega}}{\Omega}-{\gamma})sin{\theta}-{\omega}_{0}cos{\theta}]B_{\theta}+{{\tau}_{0}}^{2}B_{s}=0
\label{54}
\end{equation}
\begin{equation}
{({\Omega}r)}^{-1}[(\frac{\dot{\Omega}}{\Omega}-{\gamma})cos{\theta}-{\omega}_{0}sin{\theta}]B_{\theta}=0
\label{55}
\end{equation}
\begin{equation}
{({\Omega}r)}^{-1}B_{\theta}{{\tau}_{0}}^{2}sin{\theta}-{\gamma}B_{s}=0
\label{56}
\end{equation}
By applying the hypothesis of weak torsion ${\tau}_{0}$ , which is a
very reasonable hypothesis for astrophysical plasma and solar tubes,
reduces the three scalar dynamo equations to
\begin{equation}
{({\Omega}r)}^{-1}[(\frac{\dot{\Omega}}{\Omega}-{\gamma})sin{\theta}-{\omega}_{0}cos{\theta}]B_{\theta}=0
\label{57}
\end{equation}
\begin{equation}
{\gamma}B_{s}=0 \label{58}
\end{equation}
Though the equations (\ref{43}) and (\ref{40}) together in the
absence of twist ${\omega}_{0}=0$ yields
\begin{equation}
{\Omega}=\frac{{\Omega}_{0}}{r}e^{{\gamma}t} \label{59}
\end{equation}
and ${\gamma}=0$. Therefore, though equation (\ref{43}) yields an
apparent Lyapunov exponent and consequently a dynamo action, the
other equation (\ref{44}) shows that only a marginal dynamo is found
where the magnetic field does not grow in time, and a fast dynamo is
not possible, which is the purpose of the demonstration of
anti-fast-dynamo theorems. $\Box$ \newline Note that the constraint
of untwisting of the flux tube is not really necessary to obtain the
non-fast dynamo action of the conformally transformed thin tube, and
was only introduced to show that the Lyapunov exponent of fast
dynamo action appears into the discussion.
\newpage

\section{Conclusions} Vishik's anti-fast dynamo theorem further
generalised by Klapper and Yound is applied here to conformal
Riemann metric of thin flux tubes to show that actually even when
the tube is allowed to evolve in time no fast dynamo action is
obtained under the non-stretching constraint. To recognize when a
map or magnetic flow does represent a dynamo or non-dynamo
represents an important issue in plasma physics and astrophysics and
deserves further investigation. Earlier the role of the curvature,
either Gaussian or Riemannian, on the existence of fast dynamos was
consider by Chicone-Latushkin \cite{10} expression for ${\gamma}$ in
the fast dynamo obtained from geodesic flows in Anosov spaces of
negative constant Riemannian curvature
\begin{equation}
{\gamma}=\frac{1}{2}[-{\eta}(1+{\kappa}^{2})+\sqrt{{\eta}^{2}(1-{\kappa}^{2})^{2}-4{\kappa}}]\label{60}
\end{equation}
where ${\kappa}$ is the Gaussian curvature. This expression shows
that when this Gaussian curvature of the two-dimensional manifold
vanishes, the expression for ${\gamma}$ vanishes as well. This shows
that in the absence of curvature a marginal non-fast dynamo is
obtained. However seems that Chiconne-Latushkin result does not
applies to the conformal Riemannian non-stretched tubes discussed
here. Curvature in certain sense seems to be more connected to
folding dynamo processing that to stretching itself. Actually in
cases investigated in this paper, stretching is deeply connected to
curvature through term $K(r,s)$ in Ricca's metric. This lead us to
conclude that the curvature is actually connected to stretching and
folding. Since in the case considering in this section, twist,
curvature and stretching are almost neglected the result seems to be
physically correct. The theorems discussed here may be extended to
include diffusive plasma flows, for example. Note that since,
according to Parker \cite{15}, the tube twisting increases by radial
or cross-section expansion, the above conformal factor
$\frac{{{\Omega}_{0}}^{2}}{r^{2}}$, is compatible with the
untwisting hypothesis used in the proofs of above anti-fast-dynamo
theorems. This happens because when the radial coordinate
increases,the conformal factor acts as "damping" for this action
making the tube decreases actually this cross-section as a
compression and giving rise to tube untwisting. Another important
observation done by Friedlander and Vishik \cite{16}, tells us that
the simple geometries of magnetic field lines in steady flows it
seems unlikely that fast dynamo action could take place. A similar
statement has earlier been done by Arnold et al \cite{9}, which
showed that magnetic Reynolds number ${R^{-1}}_{m}=0$ steady flows,
only give rise to non-exponential growth of magnetic field and
therefore to no fast dynamo action. One notes that is more difficult
to handle the dynamo problem in non-Euclidean spaces, since in this
case, the frames evolve dynamically in time as well. Of course
introduction of turbulence and mean field magnetohydrodynamics
\cite{17}, introduce complications in both cases. Vishik's has also
shown that the fast dynamo action can be proved for smooth flows in
stretching and unstretching directions. Following this idea, in this
paper one has shown that it is possible to supress the stretching
directions by the action of the conformal map. In our example, the
flux tube toroidal direction is non-stretching while, the poloidal
stretching directions are suppressed by conformal map, yielding a
anti-fast-dynamo theorem. Fast conformal dynamos in periodic flows
in diffusion plasmas, and in the case of zero resistivity plasmas,
showing that unfolding Riemann-flat manifolds favors the dynamo
plasmas in flux tube while the non-stretching, untwisting tubes
favors non-dynamos or marginal ones, or yet in the sense of
anti-fast dynamo theorems a non-fast dynamo. Torus fast plasma
dynamos seems to be less favoroured, than the plasma dynamo cylin
ders \cite{18}. Jean Luc Thiffeault \cite{19} has recently called my
attention to a paper where The Riemann curvature tensor also
vanishes as a constraint to two dimensional diffusion in Riemannian
manifolds. More realistic fast dynamo picture than the one given
here, was recently given by Gilbert \cite{20} which considers
chaotic stretch-shearing flows with strong folding and cat maps in
torus as well. Strong folding Riemann curvature dynamos could be
investigated elsewhere.
\section{Acknowledgements} I am deeply greatful to Andrew Soward, Dimitry Sokoloff, Jean Luc Thiffeault, Andrew D. Gilbert
and Renzo Ricca for their extremely kind attention and discussions on
the subject of this paper. I would like to dedicate this paper to
the $65th$ anniversary of Professor Andrew Soward, which did so much
for our understanding of dynamo theory. Thanks are also due to I
thank financial supports from Universidade do Estado do Rio de
Janeiro (UERJ) and CNPq (Brazilian Ministry of Science and
Technology).

  \end{document}